\documentclass[12pt]{article}

\usepackage{enumerate}
\usepackage{amssymb,amsmath,amsfonts,amsthm,textcomp}
\usepackage{eucal,graphics}
\usepackage{endnotes}
\usepackage{natbib} 
\usepackage{setspace}

\newcommand{\e}{\ensuremath{\mathbb{E}}}

\newcommand{\p}{\ensuremath{\mathbb{P}}}
\newcommand{\q}{\ensuremath{\mathbb{Q}}}

\newcommand{\spa} {\ensuremath{\mbox{ }}}

\newcommand\ent[2]{ \citep[#2]{#1}}

\begin{document}

\title{Reciprocity as the foundation of Financial Economics\footnote{Research supported by
EPSRC grant \, EP/C508882/1}\ 
\footnote{This is an Author's Original Manuscript of an article submitted for consideration in the   [copyright ].}
}

\author{
{\sc Timothy C.\,Johnson\footnote{Maxwell Institute for
Mathematical Sciences and Department of Actuarial Mathematics
and Statistics, Heriot-Watt University, Edinburgh EH14 4AS,
UK, \texttt{t.c.johnson@hw.ac.uk}}}}

\doublespacing
\maketitle

\begin{abstract}
This paper argues that the fundamental principle of contemporary financial economics is balanced reciprocity,  not the principle of utility maximisation that is important in economics more generally.  The argument is developed by analysing the mathematical Fundamental Theory of Asset Pricing with reference to  the emergence of mathematical probability in the seventeenth century in the context of the ethical assessment of commercial contracts.  This analysis is undertaken within a framework of Pragmatic philosophy and Virtue Ethics.  The purpose of the paper is to mitigate future financial crises by reorienting financial economics to emphasise the objectives of market stability and social cohesion rather than individual utility maximisation.
\end{abstract}

\section{Introduction}

The processes of structured finance employing `special purpose vehicles/structured investment vehicles' (SPVs / SIVs) to fund  activities, whether by industrial corporations, like Enron; governments, such as the UK government's Private Finance Initiative; or  banks, through asset backed securities,  has had a significant impact on modern business practice.  However, along side its growth, structured finance has become associated with financial crises, particularly the Financial Crisis of 2007--2009 (\cite{CDK_MSS}, \cite{B_IFCEP},  \cite{AR_SCRFS}).   Recurring themes associate  the Crisis with financial innovation generating complex risks (e.g. \cite[p 55]{D_SLMF},  \cite[p 564]{C_SCGFC}, \cite[p 769]{L_CEC}, \cite[p 14]{TR})  and the inability of the mathematical techniques employed by financial economics to address these issues (e.g. \cite{B_ENSR}, \cite{D_SLMF}, \cite{C_SCGFC},  \cite[p 14]{TR}, \cite{L_CEC}, \cite{Cea_FCSFEP}, \cite{HM_SRBE}, \cite{E_FS}).  Amongst those who identify a problem with mathematics, there are two classes: those who can see a possible solution in mathematics (e.g. \cite{B_ENSR}, \cite{Cea_FCSFEP}, \cite{HM_SRBE})  and those who don't and advocate tighter market regulation (e.g. \cite{D_SLMF}, \cite{C_SCGFC},  \cite{L_CEC},  \cite{TR}).

While SPVs are frequently presented as twentieth century innovations, their fundamental character of transforming an uncertain future cashflow into a fixed and insured cashflow was exhibited in the \emph{Contractus Trinus} (the `triple' or `German' contract).  \citet[Ch 10]{N_SAU} and  \cite{D_IDCC} have given detailed accounts of the contract and the scholastic debate about their legitimacy which eventually led to their condemnation of usury in 1586 (\emph{Detestabilia avaritia}).  Securitisation and collateralisation manifested themselves as the corpo/sopracorpo structures that emerged in the thirteenth century (e.g. \citet[p 554]{P_EMFE}) and the trading of securitised assets has been widespread since the eighteenth century, particularly through the reinsurance markets (e.g. \citet[p 38]{L_FF}).

Mathematics is frequently presented as being an alien encroaching into economics (e.g. \cite{W_HEBMS}).  However the foundation of European mathematics is in Fibonacci's 1202 \emph{Liber Abaci}, a text for merchants \citep{LA}, and the subsequent  mathematisation of the physical sciences was stimulated by the mathematical analysis of financial phenomena (e.g. \cite{H_OSM}, \cite{K_ENFC}).  The    emergence of mathematical probability, at the heart of all science, is out of the ethical analysis of commercial contracts (e.g.  \cite{F_SC}, \cite{S_BECM}, \cite{B_DLLA},  \cite{S_CAT}). Rather than mathematics encroaching on economics, the historical account suggests that economics has generated mathematics that is then used in other domains.   The issues that many of the critics of the contemporary use of mathematics in economics raise are not relevant to mathematics, as conducted throughout history, but to modern mathematics.  The nature of mathematics has changed in the twentieth century;  pre-modern mathematics developed in the vernacular of financial practice and  between the eighteenth and twentieth centuries was motivated by observation of natural phenomena.  In the second half of twentieth century a theoretical, formalist-deductivist, approach dominated mathematics, and this was  adopted by orthodox economics.  While this approach to mathematics has been rejected in the physical sciences \citep{G-M_NCH} it seems to persist in economics (e.g. \cite{W_HEBMS}, \citet[Ch 10]{L_RE}).

This sketch suggests that while financial technology has not changed significantly over the centuries, attitudes to   mathematical practice have.  Mathematical approaches to financial problems before the nineteenth century were explicitly ethical.  Authoritative assessments of the causes of The Crisis that have looked beyond the field of economics have focussed on the ethical nature of the financial failures. The {Financial} {Crisis} {Inquiry} Commission (FCIC) concluded that in the lead up to The Crisis    there had been a  ``systemic breakdown in accountability and ethics'' \citep{FCIC}.    The UK's Parliamentary Commission on Banking Standards \citep{PCBS} pointedly titled their comprehensive report of 2013 ``Changing Banking for Good'', emphasising that the direction of change should be in an explicitly moral direction.  The argument in both reports is that improving ethical behaviour is a key component of improving financial stability.  The role of mathematics is  peripheral but significant in that it legitimised financial innovation by creating a false sense of confidence (e.g. \citet[p 44]{FCIC}, \citet[vol. 2, para. 60]{PCBS}): mathematics speaks with an indubitable authority that need not be questioned \citep[Ch 10]{L_RE}.

In light of these observations the purpose of this paper to investigate links between contemporary financial economics  and ethics in order to make a contribution to mitigating financial crises.  This  theme has been addressed by a variety of authors, for example,   Kevin Jackson \citep{J_SBFC} tackles it tangentially by addressing failures in the curricula of Business Schools, an issued examined in detail by Jason West \citep{W_EQF}. Pre-dating the events of 2007,  James Horrigan, George Frankfurter and Elton McGoun have examined the underlying  ideology of  financial economics (\cite{H_ENF}, \cite{FMcG_IIIIFE}, \cite{F_TFM}).  

The relative paucity  of literature on ethics in financial economics, as compared to scholarship on ethics in other technology based professions,  probably comes about because the discipline is explicitly mathematical and mathematics strives to be infallible by maintaining a strict fact/value dichotomy. Therefore, in order to achieve its objective the paper will abandon the fact/value dichotomy in financial economics.  This path is justified by adopting a Pragmatic approach, which is characterised by arguing that knowledge has no certain, infallible, foundations and because scientists are part of the system they observe, there can be no real distinction between what is and what ought to be.  A consequence of adopting a Pragmatic approach is that the historical evolution of beliefs and the role that practice has in generating theory are an important themes running through the paper. \citep[Introduction]{M_INP}

Pragmatism addresses the question of `what is'; the `fact', the paper addresses `what ought to be'; the `value', by taking a Virtue Ethics approach, as employed in the seventeenth and eighteenth centuries and is a growing area of interest (e.g. \cite{FSB_PVIP}).  Taken together, these two approaches distinguish the paper from either Horrigan or   Frankfurter and  McGoun.

As a consequence of taking this  approach is that  this paper  concludes that contemporary financial economics has an implicit foundation in ethics, specifically Justice expressed as balanced reciprocity.  This conclusion is arrived at by showing that one of the fundamental theory of financial economics, the Fundamental Theorem of Asset Pricing (hereafter `FTAP'), has its basis in the virtue  `Justice'.  The  FTAP is  the mathematical theory underpinning modelling frameworks such as  Black-Scholes-Merton, Cox-Ross-Rubinstein, Heath-Jarrow-Morton and the LIBOR Market Models and is the central theory in contemporary mathematical approaches to pricing derivatives.  Its significance is in unifying various strands in financial economics: Samuelson and Merton's use of  stochastic calculus; CAPM, developed by Treynor and Sharpe;  martingales, employed by Fama in the development of the Efficient Markets Hypothesis; Arrow and Debreu's concept  of incomplete markets.  In accomplishing this unification it represents a Kuhnian paradigm for financial economics. 

Arguing that reciprocity is implicit and embedded in contemporary financial economics is unorthodox; it is more normal to argue that markets are socially destructive (e.g. \cite{K_SLMF}, \cite{R_MFEM} ) or have the potential for corruption if not constrained (e.g. \citet[p 55]{D_SLMF},  \citet[p 564]{C_SCGFC}, \citet[p 769]{L_CEC}, \citet[p 14]{TR}), and so financial economics is immoral in facilitating this corruption.  These attitudes have a powerful influence on how investigations of the causes of The Crises are framed. In order to address these framing issues, we discus the changing social attitudes to markets  in order to provide some context to the thesis that financial economics is implicitly ethical.

The paper is presented as follows.  It begins with a general introduction to Pragmatism, in the context of other epistemological frameworks, and Virtue Ethics, in the context of other ethical frameworks, before moving onto an overview of cultural attitudes to the morality of markets.    Section \ref{s2} describes the emergence of probability between the thirteenth and seventeenth centuries.  It is significant as discussions of ethics and financial economics  rarely  consider the interaction between theory, practice and morality  before Adam Smith.  This is an important oversight as the ethical assessment of medieval and seventeenth century commercial practice had a profound effect on the subsequent development of science, in particular of mathematical probability.  These arguments provide a basis for a deconstruction of the FTAP as a theory based on reciprocity, which is undertaken in Section \ref{s4}.  The final Section extends the observation that the FTAP is based on reciprocity to the conjecture that the whole of financial economics is founded on the norm.  This develops into a discussion of the relationship between commerce and Pragmatic philosophy and offers a novel explanation for recent financial crises as well as a possible regulatory approach to mitigate future crises.

\section{Fact, Value and the Morality of Markets}\label{s1}

\subsection{Fact}
Determining what \emph{is} is part of epistemology, and epistemological theories can be broadly separated into two classes, \emph{foundational} and \emph{coherentist}.  ``Foundational theories attempt to ground knowledge in a solid base such as sense experience [Empiricism] or a priori reasoning [Realism]. In contrast, coherentists  argue that there are no foundations for our beliefs, whose justification derives from how well they fit together with each other.''\citep{TB_EMNP}

Realism argues that there is an `intelligible' universe, of immutable truths,   and a `sensible' universe, that is actually experienced and undergoes change.  `Truth' transcends  experience and can be  established only through abstract thinking (Rationality).  Realism, in the Christian and Islamic traditions, can be traced to Plato's Theory of Forms (or Ideals) and was central to Descartes' and Kant's philosophy. In this framework,  there is a hierarchy of knowledge with mathematics being closer to  `truth' than experimentation. Beliefs in the immutability and   indubitability  of mathematics became embedded in western philosophy with the Neo-Platonists, such as   Augustine of Hippo, who associated mathematics with a transcendental deity \citep[p 46]{A_FCW}.

Contemporary Empiricism argues that there are two types of truth: tautologies, established through formal mathematics or logic; and factual statements that can be verified by employing the `scientific method' to guide observation and analysis.  A principal of Empiricism is that while `Truth' might be unachievable, the scientific method will converge towards a close approximation of true facts.  European Empiricism has its roots in Greek Epicureanism and became dominant in British philosophy through Francis Bacon, John Locke, David Hume and J. S. Mill.  Logical Positivism, a form of Empiricism,  emerged in Vienna in the early twentieth century and became significant in North America in the 1940s. 

While   Empiricists reject the metaphysics of Realism, they generally do not challenge the status of mathematics and  created a special class of Truth related to  Hilbert's Formalism.  To appreciate the distinction between Realism and Empiricism, a Realist might claim that ``$2+2=4$ was true at the time of the dinosaurs'', implying the mathematics is independent of human thought (\emph{synthetic a priori}); an Empiricist would claim the statement is a tautology: $2:=1+1, 4:=1+1+1+1$ and so $2+2=(1+1)+(1+1)=4$ (\emph{analytic a posterior}).

Within economics, Realism is associated with Neo-classical theories that employ equilibrium and rationality and resort to \emph{ceteris paribus} arguments to explain why economic facts (what is experienced) rarely conform to Neo-classical theory \citep{DM-R_MP}. Contemporary economics in the Empirical spirit includes experimental and behavioural economics (e.g. Vernon Smith, Daniel Kahneman).

Realism and Empiricism are foundational theories  built on the idea that the scientist is a `spectator' observing phenomena at a distance and as a result `Truth' is a static representation of  phenomena. Pragmatism, on the other hand, argues that the scientist is part of the phenomena they observe and so `objectivity' is unachievable, as a result `Truth' is just what competent, rational enquiry  produces and will evolve in time. Pragmatism emerged in the late nineteenth century with Charles  Peirce, William James and John Dewey and then  by Willard Van Orman Quine, Hilary Putnam and Richard Rorty,  and  more recently  by Susan Haack, Robert Brandom and Cheryl Misak,  amongst others.  While closely associated with American philosophy, there have been Pragmatic strands in French thought,  associated with Greek Sophism, notably the `occasional Pragmatism' of Henri Poincar\'{e} \citep{H_HPTPS} while  \'{E}mile Durkheim acknowledged the usefulness of Pragmatism in destroying ``the cult of truth'' (\cite{L_NRE}, \citet[pp 69-72]{D_PS}). In Britain, Pragmatism has been associated with the Cambridge School, particularly with  (the later) Wittgenstein and now with Huw Price.  Pragmatism overlaps Empiricism (e.g.  Peirce,  Quine)  and Realism (e.g.  Brandom,  Putnam) \citep[pp xvi--xvii]{R_CP}.

%There are two metaphors discussed by Thagard and  Beam \citep{TB_EMNP} that are useful when approaching Pragmatism.   Peirce challenged classical Realism by arguing that `reasoning' was not a chain of deductions that is reliant on the `weakest link' holding, but rather a body of knowledge is like a cable with each fibre of the cable representing a belief: if a fibre fails the cable remains in place.  Otto Neurath  challenged Empiricism by presenting a body of knowledge as a ship and scientists (sailors) must maintain the ship (revise beliefs) at sea, they do not have the opportunity to completely rebuild the ship. 

Pragmatism does not assign a special status to mathematics, in the way that Realism and Empiricism do,   Putnam argues that
\begin{quote}
we learn what mathematical truth is by learning the practices and standards of mathematics itself, including the practices of \emph{applying} mathematics. \citep[p 66]{P_EWO}
\end{quote} 
while Poincar\'{e} observed that
\begin{quote}
The principal aim of mathematical education is to develop certain faculties of the mind, and among these intuition\endnote{The definition  of `intuition' here can be read as `The action of looking upon or into; contemplation; inspection; a sight or view.' (OED 1), from the Latin \emph{intuitus}: to look, rather than the philosophical definition `The immediate apprehension of an object by the mind without the intervention of any reasoning process' (OED 3). }  is not the least precious. It is through it that the mathematical world remains in touch with the real world, and even if pure mathematics could do without it, we should still have to have recourse to it to fill up the gulf that separates the symbol from reality. \citep[p 449]{P_SM} 
\end{quote}

Pragmatism is being associated with a revival of `classical economics' \citep{M_RCPE}, is close to Pasinetti's description of the  Cambridge School of Keynesian Economics \citep{P_CSKE}, relates to Deirdre McCloskey's economics founded on rhetoric (discourse) \citep{M_BD}, has been linked to behavioural economics \citep{K_DPEM} and institutional economics \citep{B_PEWJC}.  Pragmatic approaches are distinguished by acknowledging ethical features of economic behaviour and emphasising the role of uncertainty (\cite {J_WTB}, \cite {D_QC}).  For example, Friedman's argument in \emph{The Methodology of Positive Economics} appears to share principles of Pragmatism \citep[p 2]{K_DPEM}: ``[Positive economics'] performance is to be judged by the precision, scope, and conformity with experience'' \citep[p 4]{F_MPE}.  However, Friedman's rejection of a normative, subjective,  dimension to economics and a faith in the ability to verify stable economic theories makes it incompatible with Pragmatism.

\subsection{Value}

%Putnam criticises  the Fact/Value Dichotomy \citep{P_CFVD} with the observation that when  Empiricists insist on the distinction, they are exclusive in associating values with ethics.  Concepts such as `coherence', `plausibility', `simplicity', and  so forth, are also values that Empiricists implicitly regard  as objective, raising the question  whether ethical values are also objective.  Central to this observation is the belief that values are embedded in scientific processes; they are so much part of life that we cannot avoid them.

Ethical frameworks, in the Western tradition, are usually classed as being  Deontological, Consequentialist or Virtuous.  Deontology can be typified as ``Thou shalt / shalt not'' and guides \emph{action} on the basis of laws, rules or principles. Since an individual cannot be subject to a law unless it has been promulgated,  Deontology is linked to with philosophical systems that are based on `divine' or `natural' law, such as Realism and Stoicism \citep[p 14]{A_MMP}.  The practical problem with Deontological Ethics is that basic rules such as ``Thou shalt not kill'' have caveats while other prohibitions become redundant, or need revising, as society evolves.  In the context of contemporary economics, Deontological Ethics has been employed  in financial regulation (e.g. Pillars I \& II of Basel II)  and has been  criticised for being over-bureaucratic and rigid  while susceptible to `gaming'; adhering to the letter of the law but not the spirit. \citep[pp 23--26]{vS_BUD}

Consequentialism  attempts to judge the value of an action in terms of its \emph{consequences}.  This approach has its roots in ancient Chinese Mohism and Greek Epicureanism, developed in opposition to Platonism and Stoicism.  The approach became fully developed in the nineteenth century  with a trio of British philosophers, Bentham, Mill and Sedgewick, who  argued that one should ``Act always in such a way as to promote the greatest happiness to the greatest number''.  

On the basis of Consequentialism and David Hume's distinction of `what is' and `what ought to be',   `value--neutrality' was established in economics: since we  have `objective access' to the empirical world' and are `rational beings', we are able to calculate the consequences of our economic actions \citep{WH_EVET}.  A problem with this value-neutrality, described by Robert Heilbroner, is that it misses the critical fact that 
\begin{quote}
 the objects observed by the social scientist all possess an attribute that is lacking in the objects of natural universe.  This is the attribute of consciousness -- of cognition, of ``calculation'', of volition \citep[p 133]{H_EVFS}
\end{quote}
The importance of `volition' had been  recognised by   Oskar Morgenstern, who  objected to perfect foresight based on calculation because
\begin{quote}
always there is exhibited an endless chain of reciprocally conjectural reactions and counter-reactions.  This chain can never be broken by an act of knowledge but always through an arbitrary act -- a resolution. \citep[quoting Mogernstern on p 129]{M_WWNMTA}
\end{quote}
%Practically this means  that while  we could   incorporate the possibility of a Japanese earthquake into the modelling of asset prices,   it would be impossible to account for the behaviour of Nick Leeson in destroying Barings' Bank.  

As well as questioning the basic ability to predict in a social context,   Consequentialism has been criticised  because obviously immoral acts, such as the execution of the innocent, could be justified either by the hope of good consequences or the fear of bad \citep[p 14]{A_MMP}.  In response to the problems of Deontological and Consequentialist Ethics, many  argue that  ethics should focus on the judgement of the \emph{agent} taking the action that has consequences, or Virtue Ethics.

Virtue Ethics, in the Western tradition, is associated with Aristotle, in particular \emph{Nicomachean Ethics} in which virtues are the ``characteristics that enable individuals to live well in communities'' \citep[p 247]{P_CP}.  Aristotle's ethics do not distinguish reason and emotion, as Hume did in the eighteenth century, nor do they define absolute standards, rather Virtue is a consequence of personal reflection \citep[pp 6--8]{vS_VEAP}.  This opens Virtue Ethics to the criticism that it cannot be codified into a set of rules that any person could apply to determine ethical action in any situation.  However, this criticism  assumes such a reduction is possible, and implicit in this is that the environment is stable and predictable.  The advantage of Virtue Ethics is precisely that it can accommodate unforeseen circumstances. 

%However,  Virtue Ethics is  open to the criticism of Relativism.  For example, Abelard, who was the first western European to highlight the tripartite classification of ethics in \emph{Dialogue of the Philosopher with the Jew and the Christian}, was condemned as a heretic on the basis that he argued that those who were responsible for the crucifixion of Jesus  were not necessarily doing wrong if they believed they were fulfilling their social obligations \citep[p 53]{L_MT}. 

%Medieval Scholars approached Virtue Ethics using the same framework that they used to study physics or  medicine, by  blending   elements, or humours, in the right manner.  They identified seven elements of morality,  the four `Cardinal' virtues; Courage; Justice; Temperance; and Prudence,  and three, so-called, `Christian' virtues:  Faith;  Hope; and Charity.    For example, blending (tempering)  Justice, Courage and Faith result in honesty \citep[p 361]{M_BV}.  An ethical life was one that exhibited all, not just some, of the virtues and anyone, priest prince or merchant, was virtuous providing they got the balance right.  

Medieval Catholic Scholars approached Virtue Ethics using the same framework that they used to study physics or  medicine, by  blending   elements, or humours, in the right manner. Their elements of morality are the four `Cardinal' virtues; Courage; Justice; Temperance; and Prudence,  and three, so-called, `Christian' virtues:  Faith;  Hope; and Charity.  All these virtues existed in pre-Christian Greek and Roman philosophy, which influenced both Judaic and Islamic thought.   Chinese (e.g. \citep{D_SECVE}) and Indian philosophy  both have  versions of Virtue Ethics that can be mapped onto the European framework. Particularly relevant is  the first century Mahayana Buddhist \emph{Vimalakirti Sutra} that tells the story of how a virtuous merchant instructs both  kings and monks.

While it is conventional to associate  Deontology  with Realism and Consequentialism with Empiricism, associating Virtue Ethics with Pragmatism is less conventional.  There are links, notably through John Dewey \citep{C_VEDM} and in the discussion of reciprocity \citep{P_CMSAD}.  More broadly, Aristotle argued that  excellence of character [\emph{\={e}thik\={e}}] derives from `habituation' [\emph{\v{e}thos}] \citep[1103a15--20]{BR_ANE}.  This can be related to  the technical term `Pragmatism', which  is derived from the Greek word describing `deed, act, affair, matter, business' [\emph{pragma}];  both terms emphasise `practice' over `theory'. 

With these observations in mind, Khalil  makes the point that ``true [Pragmatic] inquiry  cannot take place in an ivory tower'' \citep[p 2]{K_DPEM} and discourse, in enabling rational enquiry,  is central to Pragmatism. In particular  J\"{u}rgen Habermas has developed the principle of  `communicative action', where communities solve problems through discussion.  Habermas defines a `norm' as a ``universally valid statement of obligation'', which some might equate with a `virtue'.  The ``binding universal norm'' is `communicative action', ``norms of communication governed by the ideal of rational discourse'' and, according to Putnam,  the ideal of rational discourse is governed by 
\begin{quote}
the norm of \emph{sincerity}, the norm of \emph{truth-telling}, and the norm of asserting only what is \emph{rationally warranted} \ldots [and] is contrasted with \emph{manipulation}. \citep[pp 113-114]{P_CFVD}
\end{quote} 
Putnam observes that the problem of leading an ethical life is a fact of life that cannot be solved  \emph{ex cathedra} by philosophy external to the individual, such as with Deontology   and Consequentialism.  %Rather, ethical disagreements are resolved through `communicative action' in a social context and the norms that govern this `communicative action' have the features of `virtues' (i.e. tempering, justice, faith, charity and prudence).  However, in advocating pluralism  Pragmatism, like   Virtue Ethics, has to defend itself from the criticism of Relativism (e.g. \cite{R_CP}, \cite[p 53]{L_MT}, \cite{F_RPPS}).

\subsection{The Morality of Markets}

Albert Hirschman has provided a description of four different views on the relationship between markets and morality: \emph{doux-commerce}, self-destruction, feudal-shackles and feudal-benefits \citep{H_RIMS}.  The idea that commerce improved society was prevalent throughout the eighteenth century.  In 1704 technical text on commerce argues ``Commerce attaches [men] to one another through mutual utility''; while in \emph{The Rights of Man} (1792) Thomas  Paine writes ``[Commerce is a pacific system, operating to cordialise mankind''. In the intervening years Montesquieu, Hume, Condorcet and Adam Smith all agreed that commerce was a powerful civilising agent, promoting honesty, industriousness, probity, punctuality, and frugality, in contrast to the excesses of absolute monarchies.

Following the Industrial Revolution, these attitudes all but disappeared and were replaced by views that blamed the collapse of morality on the influence of capitalism.    Commerce was seen as commodifying human interaction, ``custom is replaced by contract'', and on this basis Romantics saw capitalism as being un-natural and undermined traditional hierarchies while  Marxists believed that commerce's alienation of the proletariat along with capitalism's instabilities would lead to revolution. Others believed that the success of capitalism, founded on frugality and probity, would be so great that society would eventually become dissolute, seeking instant gratification, echoing the rise of Republican Rome and the fall of Imperial Rome.

Both the \emph{doux} and self-destructive views of commerce represented capitalism as a powerful force driving social change.  When capitalism did not collapse, the emphasis changed and capitalism was not seen as strong but weak: the bourgeoisie  were unable to escape traditional social forces.  %Marx, while arguing that English capitalism would destroy itself,  also argued that German capitalism was hindered by antiquated social and political structures.  For Schumpeter, in \emph{The Sociology of Imperialisms} written during World War I, Weber's `spirit of capitalism'  was no where in the warmongering of the age.  Schumpeter's views contrasted to the optimism of the pre-war sociologists Durkheim and Simmel who both saw echoes the ties that bound traditional societies in contemporary commercial relations.
The United States of America, not bound by ``feudal--shackles'' seemed to have an advantage over Europe between 1914 and the sixties.  Capitalism, led by America, seemed to rediscover its confidence in solving society's problems after the Second World War.  But this confidence was lost in the economic malaise of the seventies.  Because America did not have the feudal past of Europe it did not have social and ideological diversity and so reforms, such as Roosevelt's New Deal, were vulnerable to a ``tyranny of the majority''; America missed the feudal--blessings.   Daniel Friedman has recently presented the relationship between markets and morals as a difficult marriage: ``where markets sabotaged morals, and morals hurt markets'' \citep[p 4]{F_MMEAMW}. In the aftermath of The Crisis, and of particular interest to this project,  Johan Graafland has related contemporary economic literature to the  \emph{doux-commerce} and self-destruction theses in the context of Aristotelian Virtue Ethics \citep{G_DMCOV}.

%The \emph{doux-commerce} thesis is a powerful argument in favour of markets yet rarely figures in neo-classical economics.   Hirschman, in 1982, explains this omission by arguing that the neo-classical program will not accommodate sociological considerations, but  does acknowledge that economics was changing, with the introduction of behavioural economics, in particular results such as Prisoner's Dilemma  that highlight the role of co-operation  in economic affairs.

Marion Fourcade and Kieran Healy \citep{FH_MVMS} have recently returned to Hirschman's characterisation and argue that it is still valid today, but have added a fifth characterisation: Moralized Markets.  In their paper Fourcade and Healy identify the four strands of Hirschman's thesis in recent scholarship starting with \emph{doux-commerce} summarising Deirdre McCloskey's argument that markets nurture ``bourgeois virtues'' as
\begin{quote}
Commerce teaches ethics mainly through its communicative dimension, that is, by promoting conversations among equals and exchange between strangers. \citep[p 287]{FH_MVMS}
\end{quote}
Researchers performing empirical studies on the Ultimatum Game (introduced the year of Hirschman's thesis, by \cite{GSS_EAUB}), argue that commerce fosters co-operation, particularity amongst strangers while others support Hayek's argument that ``Capitalism makes you free''.  Finally, some economists look for evidence that markets are the best motor for innovation. In opposition to these strands, economists are arguing that instead of virtue we have envy, instead of co-operation there is coercion, freedom does not equate to populism and creativity is being stifled by copyright. 

While economists seem to focus on the robust nature of markets, able to create or destroy society, sociologists tend to study the feebleness of markets.  Following Weber, some authors argue that markets are consequences of cultural legacy, of institutions.  The new  Moralized Markets thesis goes further, it characterises markets as `cultures', not simply a consequence of culture, which  ``are explicitly moral projects, saturated with normativity.'' \citep[pp 299-300]{FH_MVMS}.

Fourcade and Healy identify three strands of the  Moralized Markets thesis. Firstly, there is the view that markets have a role in creating moral boundaries, as McCloskey argues.  This approach follows Durkheim, who argued that morality is not fixed by some `Ontological' ethical standard (that is, one  fixed and derived from a single issue \citep[p 19]{P_EWO}); rather, morality is defined by the group.  

The second strand builds on the first by turning to the sociology of science, where an emphasis is placed on impartiality in evaluating scientific knowledge (i.e. it studies failures as well as successes).  A key theme in this approach is to study what Michael Callon called the `performativity of markets', that economic theory \emph{drives} economic behaviour, rather than that  theory \emph{describes} economic behaviour; in the words of Donald MacKenzie, financial economics is ``An engine not a camera''.  These views are close to the Pragmatic attitude, that Empiricism and Rationalism fail by not acknowledging that scientists are an active part of the system they observe \citep[pp 35--37, 50--53]{B_PI}.

While the second strand of the Moralized Markets thesis focuses on behaviour at the micro level, the third strand considers economic rules at the macro level and how they are saturated with normative considerations.  For example, when Friedman made the case for positive economics it was ``to make correct predictions''  \citep[p 4]{F_MPE} he ignored  the question of what determines `correct', and this  driven by mutable normative values. For example determining `correctness' has changed with the emergence of the value `efficiency' and the decline of `social cohesion'.

\section{The Emergence of Probability}\label{s2}

\subsection{Medieval Finance}
From 1000 C.E. until about 1300 C.E. there was a rapid development of the economy in Western Europe as it evolved from an agriculturally based feudal society towards a commercially based bourgeois society, initially in Italy then, in the twelfth century, in North Western Europe.  One physical manifestation of this change was the volume of coin circulating in the European economy, as the population doubled over the three hundred years, the amount of coin per person tripled. (\citet[Chapter 3 \& 4]{P_EHME}, \citet[pp 15--16]{K_ENFC}, \citet[p 72]{E_CMA})

\subsubsection{Practice}

Medieval European merchants, unlike their contemporaries in the Middle East, India or China, had to contend simultaneously with   prohibitions on usury and the heterogeneity of currency.  Muslim merchants had usury prohibitions but homogeneous currency, Indian and Chinese merchants had to (sometimes) deal with heterogeneous currencies but without the centralised religious prohibitions on usury.

Usury derives from the Latin \emph{usus} meaning `use', and referred to the charging of a fee for the use of money.  Interest comes from the Latin \emph{interesse} and originated in the Roman legal codes as the compensation  paid  if  a contract was broken \citep[p 73]{HIR}.  Shortly after 1200 the theologian, Peter the Chanter, argued that ``a buyer or seller may be excused from usury if he exposes himself to the risk of receiving more or less'' \citep[pp 263--264]{F_SC} and this idea that usury was absent in the presence of risk became firmly established in the thirteenth century. 

The basic financial instrument at this time was the census  that originated when ninth century monasteries guaranteed a fixed regular income in exchange for a donation of land.   Censii developed to be written on the back of a diverse range of assets, including  a craftsman's labour, resembling modern day securitisation.  In time `structured'  contracts emerged such that a borrower would receive a lump sum secured against the future cash-flow from an asset,  \emph{rente \`{a} prix d'argent}, without necessarily relinquishing ownership of the asset (\citet[pp 75--76]{HIR},  \citet[pp 31--33]{EHFE}).  

Modern structured finance was anticipated in the  triple, or German, contract (\emph{contractus trinus}), developed to fund long distance trade.  It involved a loan to fund the venture (the first contract); the transformation of the variable return of the venture into fixed cash-flow (the second contract); and an insurance contract to guarantee the fixed payment (the third contract). In terms of contemporary finance this third contract is a  Credit Default Swap and the whole contract has the same structure of a Special Purpose Vehicle.  This contract was declared illicit by the Catholic Church in 1586 on the basis that the lender received a risk-less return. \citep[pp 209--220]{N_SAU}

The heterogeneity of currency was a consequence of feudalism and the desire of magnates to assert their authority by issuing coin.  The Italian peninsula had over twenty currencies, the Kingdom of France three, and each prince of the Holy Roman Empire would mint their own coin.   Alfred Crosby describes the activities of a Tuscan merchant in supplying cloth to Venice from Mallorcan wool that involved at least five currencies \citep[p 201]{C_MR}. William Goetzman explains that as a consequence of the multitude of currencies,  European  medieval merchants ``operated in a world of complete relativism'' \citep{G_FFR} while Crosby remarks that there was an  ``abstraction of Western merchants' scale of value''  and ``no people were more obsessed with counting and counting and counting''\citep[p 72, 74]{C_MR}.

A solution to the problem of the complexity of Medieval commerce came in Fibonacci's \emph{Liber Abaci} first published in 1202, the initiant of Financial Economics (\citep[43--47]{C_MR}, \citep[Introduction]{LA}).  It was an immediate success and a second edition was produced in 1228, a remarkable feat in an age when books were hand copied \citep[p 4]{LA}.  The text introduces Arabic/Hindu numerals and explains basic arithmetic over seven chapters.  It then presents four chapters applying the theory by presenting cases on practical commercial problems.  The text finishes with a more theoretical section on iterating to a solution of a problem.  (\citep{LA}, \citep{G_FFR}) 

Before the \emph{Liber Abaci}, European merchants, like their contemporaries across the globe, would have used an abacus  to perform arithmetic calculations, and once a calculation had been made, it was recorded.  The technologies described in the \emph{Liber Abaci}, particularly Hindu numbers, meant that merchants could write down their calculation method, the algorithm, which could be copied and modified by others. Knowledge, in the form of best practice,  could be created, distributed and improved.

\emph{Abaco} or \emph{rekoning} schools sprang up throughout Europe teaching apprentice merchants the techniques originating the \emph{Liber Abaci}.    The impact of these abaco schools  was enormous, algebra became an important tool used by the large and influential community of Europeans and would provide the reservoir of mathematicians on which the scientific developments of the seventeenth century were built.  The unique circumstances of medieval European commercial practice offer a solution to Needham's question that asks why European technological development accelerated so much faster than Chinese after 1600. (\citet[Chapter 1]{H_OSM}, \citet[Introduction]{LA}, \cite{H_AT})

\subsubsection{Theory}

The societal changes before 1200 led to a need to revitalise the Catholic Church, particularly to combat unorthodoxy  such as Catharism.  The Dominican and Franciscan orders were established to engage with the emerging bourgeoisie and would come to dominate Scholasticism, the intellectual movement that integrated Greek philosophy and Christian theology in Europe's universities until the Reformation.

The science that emerged in Western Europe in the seventeenth century is distinctive in its use of mathematics to describe the laws of nature.  The Greeks, and their Muslim successors, generally regarded `pure' mathematics as being irrelevant to the sensible  world while Chinese scientists used mathematics to calculate but not to describe (\citep[p 16]{C_MR}, \citet[p 164]{D_RS}, \citep[p 53]{F_SFTHY}).  Richard Hadden, Alfred Crosby and Joel Kaye have all argued that the `mathematisation' of European science began with the synthesis of commercial practice and Scholastic ethics in the thirteenth and fourteenth centuries (\cite{H_OSM}, \cite{C_MR}, \cite{K_ENFC}).

A key component of this synthesis was Aristotle's   \emph{Nicomachean Ethics} that addresses  how an individual can live as part of a community and it discusses economics  in Book V  in the context  of the virtue of Justice.  Aristotle   saw reciprocity in exchange as being important in binding society together, and Aristotle believed  exchange was performed  to correct for inequalities in endowment and to establish a social equilibrium, not in order to generate a profit (\citet[p 51]{K_ENFC}, \citet[1133a15--30]{BR_ANE}).  

Aristotle distinguishes economic  justice into two classes, distributive and directive (or corrective, restorative).  Distributive justice is concerned with the distribution of common goods by a central authority in proportion to the recipients' worth and is determined by equating  Geometric Proportions.  Directive justice  applies in cases where the parties are considered to be equal, for example in commerce, in which case justice is determined by equating Arithmetic Proportion and is based on reciprocity (\citet[p 41--43]{K_ENFC}, \citet[1130b30--31a5]{BR_ANE}).

What is most striking in Aristotle's treatment of economic exchange is that he approached it as a mathematical problem.  
This is remarkable in itself because Aristotle rarely applied mathematics to the sensible world elsewhere (\citet[p 75]{H_OSM}, \citet[p 13]{C_MR}, \citet[1094b15--28]{BR_ANE}).  
 Aristotle realised that if there was to be equality and Justice then 
\begin{quote}
everything that is exchanged must be somehow comparable.  This is the role that is fulfilled by currency [\emph{nomisma}], so that it becomes, in a way, an intermediate \citep[1133a19--20]{BR_ANE}
\end{quote}  
These lines are significant for two reasons. Firstly the word \emph{nomisma} for currency/money  is related to the concepts of custom and law, not to `labour and expenses'.  Secondly,  `intermediate' is in the sense of a mediator between two objects, rather than simply as a token, which is a more modern interpretation.  Furthermore,  Aristotle defined the quality that money measured by the word \emph{chreia}, which was initially translated to \emph{opus} (work), but was later corrected to  \emph{indigentia} (need) \citep[pp 68--70]{K_ENFC}.  This is important because it demonstrates that Aristotle and the Scholastics viewed money as a social construction binding society by allowing an exchange based on need, rather than as a simple commodity facilitating the exchange of sensible quantities, such as labour and expenses.

The significance of the Scholastic analysis to the development of science was that when Aristotle discussed measurement in the context of physics he argued that the measure shared the `substance' of the measured; this meant that wine was incommensurable with cloth, time incommensurable with space. The Scholastics realised that  money was a very special measure; it applied to all goods in a market, and only occasionally shared the substance of the goods.   This insight enabled them to revolutionise the concept of measurement, in a way that contemporary Muslim scholars did not, and allowed   Jean Buridan to identify the concept of inertia. (\citet[p 263-268]{B_HM}, \citet[p 67--74]{C_MR}, \citet[pp 65--70]{K_ENFC})

Out of Aristotle's discussion of market exchange, Scholastics developed the concept  of the `Just Price', which has been the subject of considerable modern debate.  For example, Raymond de Roover \citep{dR_CJP}, argues against viewing the Just Price in a Marxist, labour theory of value, sense but rather as the market price, in a neo-classical, liberal sense.  However, neither of these modern positions corresponds to how the Scholastics viewed the concept.  The interpretation of the Just Price we shall employ, based on the Scholastic attitudes to Aristotle's description of exchange,  is the one discussed by Fabio Monsalve \citep[pp 6--7]{M_SJPVCMP}.  The  Just Price represents an ``intellectual construct: an ideal price that guarantees equality in exchange'' and that it represents a mathematical `medium' or a `mean'.

Monsalve points out  that Scholastic analysis was conducted in  a definite moral frame of reference, and so the Just Price ``could not refer indiscriminately to whatever price might be obtained in the market'' \citep[p 8, quoting Langholm]{M_SJPVCMP}.  This aspect was discussed in detail by the Scholastics prompted by a question `Whether the seller is bound to state the defects of the thing sold?'  posed by the important Dominican Thomas Aquinas \citep[ II, ii, qu. 77, art. 3, ad. 4]{Summa}.   Specifically Aquinas addresses a problem originating in Stoic philosophy relating to the conduct of a merchant carrying a supply of food to a starving country.  The merchant knows that they are the first of a number of merchants bringing food, the question is, should he sell the food at the high `market' price or a lower price based on his knowledge.

Kaye  makes the point that Aquinas separates the Just Price, determined by divine law, from the `market price', established by men, and explains that if the Just Price equated with the market price then an ``individual's responsibility in economic activity is effectively eliminated'' \citep[p 98]{K_ENFC}.  Despite realising this distinction, the answer from Aquinas is a little surprising.   Aquinas observes  that the merchant  may \emph{believe} that there are more grain shipments on the way, but does not \emph{know}: the future is uncertain.  On the basis that there is no certainty, and on the authority of Peter the Chanter, the merchant may charge the going market price, making an excessive but nevertheless legitimate profit, though it would be more virtuous to charge the lower price.

Aquinas' argument was criticised by a leader of the `Spiritual Franciscans', Pierre Jean Olivi.  The Spiritual Franciscans  argued that the vow of poverty meant monks should limit their use of property,  \emph{usus pauper}, not  simply not own property.  As a consequence of this extreme position Olivi was posthumously condemned as a heretic in 1326, hindering the subsequent transmission of his thought.  The Franciscans, unlike the empirical rationalist Dominicans such as  Thomas Aquinas, were fideists and this philosophical approach  meant that Olivi argued that the metaphysical {probability} of more grain arriving  had a certain reality, which Aquinas was ignoring  \citep[p 121]{K_ENFC}.  Olivi said
\begin{quote}
 The judgement of the value of a thing in exchange seldom or never can be made except through conjecture or probable opinion, and not so precisely, or as if understood and measured by one invisible point, but rather as a fitting latitude within which the diverse judgements of men will differ in estimation \citep[p 124]{K_ENFC}.
\end{quote}
This distinction is essential in demarcating the Just Price, an imprecise abstraction,  from the market price, which is observed at a fixed point \citep[Section 3.2.1]{M_SJPVCMP}.  

Olivi  seems to have interacted with merchants  and been a close observer of markets and considered a number of aspects of commerce including the problem of usury \citep[p 265]{F_SC}. Based on the principle that a lender could charge a borrower compensation for a loss (\emph{interesse}) Olivi recognised that borrowers should compensate lenders for the `probable profit' they could earn by employing capital elsewhere.  Fair exchange was a question of restoring  `probable equivalence', not of precise equality (\citet[p 119]{K_ENFC},  \citet[pp 265--267]{F_SC}).
As part of this argument Olivi  commented  that  a valuation did not only depend on `need' but also on a good's  scarcity, usefulness and desirability.  Since both need and desirability are subjective, different people will value the same good differently and based on these ideas, Olivi was able  also to explain the `value paradox' (\citet[pp 60--61]{ETBAS}, \citet[pp 123--124]{K_ENFC}).  Ultimately, according to James Franklin, Olivi thought of probability as a trade-able entity, and so could be quantified \citep[pp 266--267]{F_SC}. 

%In 1903 Poincar\'{e} pointed that without probability   ``science would be impossible'' \citep[p 139]{P_SH} while in 1927 Bertrand Russell  wrote that ``It is important to realise the fundamental position of probability in science.'' \citep[page 301]{OoP}.  Given the status of the fact/value dichotomy during the nineteenth and twentieth century it is remarkable that the idea of quantifying probability emerged out of a synthesis of financial practice and ethical theory.

\subsection{The Science of Conjecture}

The Science of Conjecture, or Probability, is the rational method for dealing with uncertainty.  Aristotle classified events into three types: certain events determined by specific causes; probable events that usually happened; and unpredictable events, including games of chance, not amenable to science \citep[p 30]{H_HPS}.  The development of Probability over the past five hundred years has been concerned principally with reducing the scope of those events `not amenable to science' in support of the Cartesian programme to place knowledge on indubitable foundations \citep[pp 281--285]{G_RD}.

While Olivi and merchants developed the idea of probability in relation to commercial exchange  and jurists and theologians addressed questions of proof, the concept of  quantifying chance did not fully materialise until the mid-sixteenth century with Cardano's \emph{Liber de Ludo Alea}. Ian Hacking has remarked \citep[Chapter 1]{EoP} that the emergence of  the concept of absolute chance was late; however, this identification of mathematical probability in the context of finance precedes both  Descartes' introduction of absolute space (Cartesian co-ordinates) and Newton's of absolute time.

Up until the 1950s, and a re-assessment of  his work by \O{}ystein Ore \citep{O_CGS}, Cardano's contribution to probability theory had been widely ignored.  In the context  frequentist interpretations of probability, that dominated the nineteenth and early twentieth centuries, it was seen as incoherent.  More recently, David Bellhouse \citep{B_DLLA} has  re-evaluated the \emph{Liber} looking at it as a humanist philosophical text, not as a mathematical document,  based on the fact that Cardano, himself, did not list it as one of his mathematical works.  Bellhouse's hypothesis is that in the \emph{Liber} Cardano is trying to establish under what grounds  gambling can be considered ethical in the context of \emph{Nicomachean Ethics}.

Cardano  latches on to the idea that Justice  is equivalent to equality and argues that in dice games `equality' was established by counting the ways a player could win and comparing that number to the ways a player would lose.  On this basis the `chance' of winning could be deduced, and if  the stakes did not match the chances, the gamble was unjust.  Summarising his findings  he states,  ``a just gamble is one between willing and knowledgeable players'', making an explicit association between science and ethics.  Almost immediately after coming to these ethical conclusions, Cardano observes that 
\begin{quote}
 These facts contribute a great deal to understanding but hardly anything to practical play {\citep[p 58 quoting from Chapter 9 of the \textit{Liber}]{GGG}}
\end{quote}
 since they offer nothing to help forecast the outcome of the dice throw.  

Throughout the sixteenth century there was a flow of mathematical thinking from the vernacular abaco schools  into academic circles, not just in the universities that had declined in status during the Renaissance but also into the courts of autocrats. For example,  Lucca Pacioli was trained in the abaco tradition, probably by Piero della Francesca, before working for a Venetian merchant.  He then took Franciscan orders and entered the University of Perugia before ultimately joining the court of the Milanese autocrat, Lodovico Sforza, where he taught perspective to Leonardo da Vinci.  Thomas Gresham was born into an important English  commercial family around the time Pacioli died and spent much of his life manipulating the exchange rate in Antwerp on behalf of the English monarchy.  When he died he created the first chair in mathematics in England with the establishment of Gresham College, which was the foundation of the Royal Society (\citep{J_GC}, \citep{B_LTTG}). Simon Stevin also trained in the abaco tradition and worked as a merchant's clerk  then as a tax official  before moving to the  University of Leiden in 1583. Ten years later he became involved with the government of the Dutch Republic during the wars of independence against Spain.  As part of the war effort, the Dutch Mathematical School was established  to train military engineers with Stevin as Director.  Stevin's vocational syllabus attracted soldiers (such as Descartes) and merchants from across northern Europe and the fees they paid raised the status of its mathematicians, who when associated with the scholastic syllabus of  abstract arithmetic and geometry, astronomy  and music, were widely regarded as irrelevant.  Stevin wrote a number of textbooks in  French or Dutch, not in the exclusive Latin of the universities, which became read widely and emphasised the \emph{practical} usefulness of mathematics in everyday life.  His influence was profound and forced other institutions change their curricula; our use of decimal notation is due to Stevin who recognised its utility in commerce. (\cite{S_SSB}, \citet[p 121]{M_MHTL}, \citet[pp 131--132]{EHFE}, \citet[p 104]{D_RS})

%When Hadden talks about science being built `on the shoulders of merchants', Pacioli, Gresham and Stevin are just some of the individuals who laid the foundations for Galileo, Descartes and Newton \citep{H_OSM}.

One problem Cardano considered was the so-called \emph{Problem of Points} which appears in a text by  Pacioli and is based on the following situation:
\begin{quote}
 Two players, $F$ and $P$, are playing a  game based on a sequence of rounds, and each round consists of, for example, the tossing of a fair coin.  The winner of the game is the player who is the first to win 7 rounds, and they will win 80 francs.
\end{quote}
  The \emph{Problem of Points} is how the 80 francs should be split if the game is forced to end after $P$ has won 5 rounds while $F$ has won 4.

Edith Dudley  Sylla  notes that the \emph{Problem} comes from the abaco tradition of using `stories' to give examples of how to solve problems in commercial arithmetic. In this case the  \emph{Problem of Points}, the story represents the case of how the capital tied up in a business partnership should be divided  if the venture has to finish prematurely \citep{S_BECM}.  

Pacioli's solution was statistical, the pot should be split 5:4.  Cardano realised this was absurd since it would give a manifestly unfair result if the game ended after one round out of a hundred or when $F$ had 99 wins to $P$'s 90.  Cardano makes the point that the correct solution would be arrived at by considering what would happen in the future, it had to be forward-looking, in particular it had to account for  what `paths' the game would follow.  Despite this insight, Cardano's solution was still wrong, and the correct solution was provided by Pascal and Fermat in their correspondence of 1654.

The Pascal-Fermat solution to the \emph{Problem of Points} is widely regarded as the starting point of mathematical probability.  The pair (it is not known exactly who) realised that when Cardano calculated that $P$ could win the pot if the game followed the path $PP$ (i.e. $P$ wins and $P$ wins again)  this actually represented four paths, $PPPP$, $PPPF$, $PPFP$, $PPFF$, for the game.  It was the players' `choice' that the game ended after $PP$, not a feature of the game itself and this represents  an early example of mathematicians disentangling  behaviour from problem structure.  Calculating the proportion of winning paths would come down to using the Arithmetic, or Pascal's, Triangle -- the Binomial distribution.  Essentially,  Pascal and Fermat established what would today be recognised as the Cox-Ross-Rubenstein formula \citep{CRR_OP} for pricing a digital call option.

The Pascal--Fermat correspondence was private, the first textbook on probability was written by Christiaan Huygens in 1656.  Huygens had visited Paris in late 1655 and had been told of the \emph{Problem of Points}, but not of its solution (\citet[p 111]{GGG}, \citet[p 67]{H_HPS}), and on his return to the Netherlands he solved the problem for himself and produced the first treatise on mathematical probability, \emph{Van Rekeningh in Speelen van Geluck} (`On the Reckoning of Games of Chance')  in 1657.

In \emph{Van Rekeningh} Huygens starts with, what is essentially, an axiom, 
\begin{quote}
I take as fundamental for such [fair] games that the chance to gain something is worth so much that, if one had it, one could get the same in a fair game, that is a game in which nobody stands to lose.\ent{H_HPS}{p 69}
\end{quote}
Probability is defined by equating future gain with present value in the context of `fair' games.

In the 1670's probability theory developed in the context of Louis XIV's \emph{appartements du roi}, thrice weekly gambling events that have been described as a `symbolic activity' not unlike \emph{potlach} ceremonies that bind primitive communities  \citep[pp 31-42]{K_ESC}.  This mathematical analysis of an  important social activity stimulated the publication of books describing  objective, or frequentist,  probability.  The Empirical frequentist approach   began to dominate the mathematical treatment of probability following the claimed `defeat', or `taming',  of chance by mathematics  with the publication of Montmort's \emph{Essay d'Analyse sur les Jeux de Hazard} (`Analytical Essay on  Games of Chance') of 1708  and De Moivre's \emph{De Mensura Sortis} (`The Measurement of Chance'), of 1711 developed in \emph{The Doctrine of Chances} of 1718 \citep{B_BF}. These texts were developed more in the context of gaming rather than in the analysis of commercial contracts and \emph{The Doctrine} was the more influential, introducing the Central Limit Theorem, and by 1735 it was believed that there was no longer a class of events that were `unpredictable'  \citep{B_BF}.  

Around 1684 James Bernoulli had begun working on problems in probability and between 1700 and his death in 1705 he worked on  \emph{Ars Conjectandi} (`The Art of Conjecturing'), a title that emphases  the practical rather than theoretical nature of conjecture, which was published posthumously in 1713.   The \emph{Ars} is made up of four parts, a commentary on Huygens' \emph{Van Rekeningh}, original work on calculating permutations and combinations, applications of these ideas to games of chance and finally the application of the ideas to ``civil, moral and economic affairs'' \citep[p 224]{H_HPS}. 

While the first three sections of the \emph{Ars}  are un-controversial, the final section is both the most significant and has proved problematic.  Bernoulli, having discussed objective probability at length introduces the epistemic, or subjective, definition of probability as ``a degree of certainty''.  Anders Hald notes that this is ``revolutionary'' because Bernoulli is applying mathematics to propositions, not just to events \citep[p 225]{H_HPS}.  This section of the \emph{Ars} is significant in that it introduces what would become known as the `Law of Large Numbers', which can be summarised as collecting a large amount of data will improve the accuracy of an observation -- providing the system was stationary \citep[p 225]{H_HPS}.  The section is problematic  because Bernoulli considered situations where the sum of probabilities could be greater than one \citep[p 27]{S_CAT}.  This is impossible if probability is calculated as relative frequency.

Sylla  compared Bernoulli's work to that of  Huygens' and other contemporaries, de Witt and de Moivre, in the process of translating the \emph{Ars} and  concluded that
\begin{quote}
 equity among associates or partners rather than probabilities in the sense of relative frequencies provided the foundation for the earliest mathematical probability theory.\ent{S_CAT}{p 13}
\end{quote}
and that
\begin{quote}
 While traditional histories of mathematical probability start with Pierre Fermat, Pascal and Huygens because they give what are from the modern point of view correct frequentist solutions to the problems of division and expectations in games of chance \ldots the foundations of Huygens' method (\ldots) was not chance (frequentist probability), but rather \emph{sors} (expectation) in so far as it was involved in implicit contracts and the just treatment of partners.\ent{S_CAT}{p 28}
\end{quote}
In the sixteenth and seventeenth centuries the motivation for the development of probability was in the ethical analysis of commercial contracts where Justice, or balanced reciprocity, `fairness' dominated.  The later Empirical approach to probability, based on observing relative frequencies, emerged out of the simpler analysis of games of chance in the context of fixed odds.

The case that Huygens was working in the context of Virtue Ethics is enhanced by recognising  the difficulty he had in translating \emph{Van Rekeningh} into Latin \citep[pp 93--94]{EoP}.  Huygens struggled to translate the Dutch word \emph{kans} (`chance', `lot'), which would normally be translated as \emph{sors}, and eventually he, or his editor van Schooten, chose \emph{expectatio}, giving the  English term `expectation' (in the mathematical sense).  However, Huygens had considered using the Latin word \emph{spes} \citep[p 95]{EoP} which was the  term for the virtue `Hope'. In French, \emph{esp\'{e}rance} is used when referring to mathematical expectation, reflecting this debate. The  Dutch, who following Stevin's focus on teaching mathematics in the vernacular, use their own terms in mathematics, in this case the  equivalent is \emph{verwachting}: hope, promise, expectation, forecast, prognosis.

Sylla also %\textsc{makes some comments about what we would now regard as incomplete markets when she} 
observes that  \emph{The Port Royal Logic},  a significant influence on Pascal, notes that ``because the house takes part of the stakes, lotteries are manifestly unfair'' and seventeenth century mathematicians recognised a distinction between actual gambles, involving transaction costs, and idealised, frictionless,  markets, suitable for the mathematical study by academics. \citep[p 327]{S_BECM}
%\begin{quote}
% assumed that gamesters were gentlemen or --women who put up a stake to play a game and divided the resulting  pot of money at the end of the game according to the previously agreed upon rules.  There was no `house' or government that skimmed part of the players' wagers off the top
%\end{quote}
%The implication is that seventeenth century mathematicians, Pascal at least, would have seen transaction costs as being unethical and, while utopian, frictionless markets  are `fair'  and so the proper subject of study.

%It is remarkable that the development of mathematical probability was undertaken almost exclusively by Augustinians: Pascal was a Jansenist; Huygens, Bernoulli and de Moivre were Calvinists;  Montmort had been trained to be an Augustinian but renounced orders to marry.  This observation is compounded by the facts that Newton was an Arian/Anglican and Leibniz a Lutheran and neither did significant work in probability, Fermat was a Catholic living in the mixed Calvinist/Catholic city of Toulouse. As Augustinians these probabilists all believed in God's pre-destination  and omniscience and they denied the existence of randomness: events  were unpredictable because man could not understand God's intentions.  The implicit determinism of the Augustinians became a standard feature of Western science, being codified by Laplace in the 1820s (\citep{D_CPE}, \citep[pp 387-390]{R_UT}).

\section{The Fundamental Theorem of Asset Pricing}\label{s4}

The Fundamental Theorem of Asset Pricing consists of two statements, (e.g. \citep[Section 5.4]{S_SCF2}) \\
\textsc{The Fundamental Theorem of Asset Pricing}
\begin{enumerate}
 \item  \textit{A market admits no arbitrage, if and only if, the market has a martingale measure.}
 \item \textit{Every contingent claim can be hedged, if and only if, the martingale measure is unique.} 
\end{enumerate}

\subsection{The context of the FTAP}
The FTAP emerged between 1979 and 1983 (\citep{HK_MAMSM}, \citep{HP1_FT}, \citep{HP2_FT}) as Michael Harrison sought to establish a mathematical theory underpinning the  Black--Scholes--Merton (BSM) equation for pricing options, which was introduced in 1973.

In the late 1960s, Fischer Black and Myron Scholes worked as investment consultants and one of the problems the pair addressed was the valuation of `warrants', options bundled with bonds.  Black was an applied mathematician who had worked in consultancy for Jack Treynor around the time that Treynor developed his version of the Capital Asset Pricing Model (CAPM).  Scholes had studied for a doctorate under Eugene Fama looking at risk-reward in the context of efficient markets \citep{S_MS}.  Black tackled the problem of pricing warrants as   an applied mathematician: the value of the warrant would be a function of the underlying asset's price and amenable to the type of calculus that had been employed since Newton and Leibnitz.  Scholes approached the problem from a financial perspective: the risk of holding a warrant  could be removed by holding a complementary (short) position in the underlying asset, by hedging.   What Scholes did not know was how to establish the size of the hedging portfolio, but when he discussed this with Black they realised the solution was in the slope of the function relating the warrant price and asset price, a  result that had been anticipated by Thorp and Kassouf \citep[pp 130--131]{ENC}.  %On the assumption of efficient markets since the value of the hedging portfolio could be calculated the value of the warrants could be deduced on the basis that ``it should not be possible to make sure profits'' \citep{BS_73}.

Simultaneously, Robert C.  Merton, who had studied advanced engineering mathematics before becoming a student of Paul Samuelson,  was considering the problem of pricing warrants from a different perspective.   Samuelson had never accepted Markowitz's criterion of trading the expected returns of a portfolio against the variance of returns \citep{S_FATPA}, which was a foundation of CAPM and Scholes' work, so  Merton tackled the problem of valuing warrants by maximising expected utility employing the stochastic calculus that had become important in aeronautical and electronic engineering. This work was published  in 1969 (\citep{S_LPS}, \citep{M_LPS}).

Despite the fact that Black  never liked Merton's highly mathematical technique,  Scholes discussed their work with Merton  in 1970.  Merton  saw how the Black--Scholes approach of hedging could be incorporated into his own continuous time models, removing the need to incorporate an arbitrary  utility function in solving the pricing problem.   Merton showed that a portfolio made up of: a single warrant, or an option; a hedging position in the risky underlying asset; and a funding position in the riskless bank account, would offer the same, certain, return as the initial cost of the portfolio  deposited in the riskless bank account.  It seemed that both subjectivity and risk had been removed from the pricing problem.  

In October 1970 Black and Scholes submitted their work to the \emph{Journal of Political Economy} and then the \emph{Review of Economics and Statistics}, but it was  rejected without review, on the basis that there was not enough economics in it.  The paper was only published by the \emph{Journal of Political Economy}  \citep{BS_73} in 1973 after the intervention of influential  academics and  shortly after the opening of the Chicago Board Options Exchange (\citep[p 314--315]{B_AG}, \citep[pp 133--136]{ENC}).  Merton published his approach almost simultaneously \citep{RTORP}.

When BSM was being developed  option pricing was a relatively unimportant activity.  Gambling legislation in the United States meant that options were only traded on `deliverable' assets, principally agricultural commodities, and these markets were  stagnant \citep[pp 142-145]{ENC}. However, following the `Nixon Shock' of August 1971, the Bretton-Woods system of fixed exchange rates collapsed and in the aftermath, interest rates, exchange rates  and commodity prices became much more volatile.  Options, which have been a feature of financial practice since the seventeenth century, and were widely traded before the suspension of the European financial markets during the First World War \citep{ABC_OA}, re-emerged as a tool to insure against volatile asset prices.

Despite the financial rational for options, their legitimacy with regard to gambling legislation was still ambiguous. The introduction of BSM delivered a mathematical equation that defined the price of an option in terms of known parameters, making their valuation  deterministic. Trading in options could  not be  gambling, given that there was no speculation in their valuation.  Donald MacKenzie reports the view of the legal counsel to the Chicago Board of Trade at the time,  Burton Rissman
\begin{quote}
Black-Scholes was what really enabled the exchange to thrive \ldots we were faced in the late 60s -- early 70s with the issue of gambling.  That fell away, and I think Black-Scholes made it fall away.  It wasn't speculation or gambling it was efficient pricing. \citep[p 158]{ENC}
\end{quote}  
Essentially a mathematical formula transformed index options from being illegitimate gambles to deterministic investments.

Both the Black-Scholes and Merton approaches to pricing options involved heuristic arguments, they were `engineering solutions'. Harrison sought to establish a rigorous option pricing `theory' to support the range of mathematical models developed on the back of the explosion in derivatives markets \citep[pp 140--141]{ENC}.  Harrison, and his colleagues, were successful in their mission and opened finance to investigation by pure mathematicians,  such as \citep{S_UFG}, \citep{DS_FTAP1}, \citep{DS_FTAP2}, and by 2000, any mathematician working on asset pricing would  do so within the context of the FTAP.

The FTAP is not well known outside the academic field  of financial mathematics. Practitioners focus on the models that are a consequence of the Theorem while social scientists focus on the original Black-Scholes-Merton approach as an exemplar. Even before the market crash of 1987   practitioners  were  sceptical as to the validity of the prices produced by their models  (\citep[pp 409-410 ]{M_BAS}, \citep[p 248]{ENC}, \citep{HT_BS}) and   today the original Black--Scholes equation is used to measure market volatility, a proxy for uncertainty, rather than to  `price' options.

However,  the status of the Black-Scholes model as an exemplar in financial economics has been enhanced following the development of the FTAP.  Significantly, the theorem unifies  different approaches  in financial economics. The most immediate example of this synthesis was that in the course of the development of the FTAP it was observed that  a mathematical  object, the Radon-Nikodym derivative, which  is related to the stochastic calculus Merton  employed  involved the market-price of risk (Sharpe ratio), a key object in  CAPM that Black used. Without the FTAP the two approaches  are incongruous \citep[p 834]{M_EW}.  Overall, as will be discussed in full in the next section,  the FTAP brings together: Merton's approach employing stochastic calculus advocated by Samuelson; CAPM, developed by Treynor and Sharpe;  martingales, a mathematical concept employed by Fama in the development of the Efficient Markets Hypothesis; and the idea of incomplete markets, introduced by Arrow and Debreu.

The synthesis by the FTAP of  a `constellation of beliefs, values, techniques'  represented a Kuhnian paradigm for financial economics focused on the Black-Scholes-Merton approach to pricing options.  The paradigm was further strengthen by the fact that the unification was presented as emerging out of  pure mathematics and appeals to Realists who believe in the transcendence of mathematics and the existence of an Idealised economic universe.  
%To paraphrase Tait 
%\begin{quote}
% A mathematical proposition  is about a certain structure, financial markets. It refers to prices and relations among them. If it is true, it is so in virtue of a certain fact about this structure. And this fact may obtain even if we do not or cannot know that it does.  \citep[p 341]{T_TP}
%\end{quote}
%In this sense, the FTAP confirmed the `truth' of many of the core concepts of financial economics in in the late 1990s. 

\subsection{An Ethical Analysis of the FTAP}

The FTAP is a theorem of mathematics, and the use of the term `measure' in its statement  places the FTAP within the theory of probability  formulated by Andrei Kolmogorov in 1933 \citep{K_FTP}.  Kolmogorov's work took place in a context captured by Bertrand Russell, who in 1927 observed that 
\begin{quote}
It is important to realise the fundamental position of probability in science. \ldots As to what is meant by probability, opinions differ. {\citep[p 301]{OoP}}
\end{quote}
 In the 1920s the idea of randomness, as distinct from a lack of information, was becoming substantive in the physical sciences \citep[pp 147--157]{vP_CMP} because of the emergence of the Copenhagen Interpretation of quantum mechanics. In the social  sciences, Frank Knight   argued that uncertainty was the only  source of profit \citep[III.VII.1--4]{RUP} and the concept was pervading  John Maynard Keynes' economics (\cite{MR_PKEPUC}, \cite[pp 84--88]{S_KRM}).  %There is an essential difference between Knight and Keynes with respect to probability:  Knight sees the issue as being epistemic: uncertainty is a consequence of a lack of knowledge, Keynes sees it as ontological: it was logically impossible to assign probabilities to most economic phenomena \citep[ Ch XXIV, 1]{K_TP}.

Two mathematical theories of probability had become ascendant by the late 1920s. Richard von Mises (brother of the Austrian economist Ludwig) \citep{PST} attempted to lay down the axioms of classical probability   within a framework of  Empiricism, the `frequentist' or `objective' approach.  To counter--balance von Mises,  the Italian actuary Bruno de Finetti presented a  more Pragmatic approach, characterised by  his  claim that ``Probability does not exist'' because it was only an expression of the observer's view of the world.  This `subjectivist' approach was closely related to the less well-known position taken by the Pragmatist Frank Ramsey who developed an argument  against  Keynes' Realist interpretation of probability presented in the \emph{Treatise on Probability} (\cite{R_TP}, \cite{RM_PP}, \cite{D_RBKELPT}, \cite{E_RPPCT}).

Kolmogorov addressed the trichotomy of mathematical probability by generalising so that Realist, Empiricist and Pragmatist probabilities were all examples of `measures' satisfying certain axioms.  In doing this,  a random variable became  a function while an expectation was an integral:  probability became a branch of Analysis, not Statistics.    

Von Mises criticised Kolmogorov's generalised framework  as un-necessarily complex \citep[p 99]{PST} while the statistician Maurice Kendall argued that abstract measure theory failed ``to found a theory of probability as a branch of scientific method'' \citep[p 102]{K_RTP}.  More recently the physicist Edwin Jaynes champions  Leonard Savage's subjectivist Bayesianism  as having a ``deeper conceptual foundation which allows it to be extended to a wider class of applications, required by current problems of science'' \citep[p 655]{J_PT}. 

The objections to measure theoretic probability for empirical scientists can be accounted for as a lack of physicality.  Frequentist probability is based on the act of counting; subjectivist probability is based on a flow of information, which, following Claude Shannon, is now an observable entity in Empirical science.   Measure theoretic probability is based on  abstract mathematical objects unrelated to sensible phenomena. However, the generality of Kolmogorov's approach made it flexible enough to handle problems that emerged in physics and engineering during the Second World War and his approach became widely accepted after 1950 because it was practically more useful.    

In the context of  the first statement of the FTAP, a `martingale measure' is a probability measure, usually labelled $\q$, such that the (real, rather than nominal) price of an asset today, $X_0$, is the expectation,  using the martingale measure, of its (real) price in the future, ${X}_T$.  Formally,
\begin{align*}
 X_0 =\e_\q\big[ {X}_T \big].
\end{align*}
The abstract probability distribution $\q$ is defined so that this equality exists, not on any empirical information of historical prices or subjective judgement of future prices.  The only condition placed on the relationship that the martingale measure has with the `natural', or `physical', probability measures usually assigned the label $\p$, is that they agree on what is possible.%\footnote{Given an event $A$, $\p(A)\Leftrightarrow\q(A)$.}.

The term `martingale' in this context derives from doubling strategies in gambling and it was introduced into mathematics by Jean Ville in a development of von Mises work of 1939.  The idea that asset prices have the martingale property was first proposed by Benoit Mandelbrot \citep{M_FFPUMMM} in response to an early formulation of Eugene Fama's Efficient Market Hypothesis (EMH) \citep{F_BSMP}, the two concepts being combined by Fama in 1970 \citep{F_ECM}.  For Mandelbrot and Fama the key consequence of prices being martingales was that the current price  was  independent of the future price and technical analysis would not prove profitable in the long run.  In developing the EMH there was no discussion on the nature of the probability under which assets are martingales, and it is often assumed that the expectation is calculated under the natural measure.  While the FTAP employs modern terminology in the context of value-neutrality, the idea of equating a current price with a future, uncertain, payoff would have been understood by Olivi and obvious to Huygens, both working in an explicitly ethical framework.

%Mathematicians concern themselves with questions of existence and uniqueness.  The existence of a martingale measure is dependent on a lack of arbitrage opportunities in the market, its uniqueness depends on whether or not all claims contingent on asset prices, such as derivatives, can be hedged.

The other technical term in the first statement of the FTAP, arbitrage,  has long been used in financial mathematics.  In Chapter 9 of  the \emph{Liber Abaci} Fibonacci discusses `Barter of Merchandise and Similar Things', 
\begin{quote}
 20 arms  of cloth  are worth 3 Pisan pounds and 42 rolls of cotton are similarly worth 5 Pisan pounds; it is sought how many rolls of cotton will be had for 50 arms of cloth.\ent{LA}{p 180}
\end{quote}
In this case there are three commodities, arms of cloth, rolls of cotton and Pisan pounds, and Fibonacci solves the problem by having Pisan pounds `arbitrate', or `mediate' as Aristotle might say, between the other two commodities.  
Over the centuries this technique of pricing through arbitration evolved into the Law of One Price: if two assets offer identical cash flows then they must have the same price.  This was employed by Jan de Witt in 1671 when he solved the problem of pricing life annuities in terms of redeemable annuities, based on the presumption that
\begin{quote}
the real value of certain expectations or chances of objects, of different value, should be estimated by that which we can obtain from as many expectations or chances dependent on one or several equitable contracts. \citep[p 313, quoting De Witt]{S_BECM}%\textit{The Worth of Life Annuities in Proportion to Redeemable Bonds}
\end{quote}
In 1908  Vincent Bronzin published a text  which discusses pricing derivatives by `covering', or hedging, them with portfolios of options and forward contracts employing the principle of `equivalence' \citep{ZH_VB2}.  In 1965 the mathematicians, Edward Thorp and Sheen Kassouf,  combined the Law of One Price with basic techniques of calculus to identify market mis-pricing of warrant prices and in 1967 they published their methodology in a best-selling book, \emph{Beat the Market}.  

Within neo-classical economics, the Law of One Price was developed in a series of papers between 1954 and 1964 by Kenneth Arrow,  G\'{e}rard Debreu and Lionel MacKenzie in the context of general equilibrium, in particular the introduction of the  Arrow Security, which, employing the Law of One Price, could be used to price any asset \citep{A_RSOAR}.  
It was on this principle that Black and Scholes believed the value of the warrants could be deduced by employing a hedging portfolio, in introducing their work with the statement that ``it should not be possible to make sure profits'' \citep{BS_73} they were invoking the arbitrage argument, which had an eight hundred year history. 

In the context of the FTAP, `an arbitrage' has developed into  the ability to formulate a trading strategy such that the probability,  under a natural or martingale measure, of a loss is zero, but the probability of a positive profit is not.  This definition is important following Hardie's criticism of the way the term is applied loosely in economic sociology, and elsewhere \citep{H_SA}.  The important point of this definition is that, unlike Hardie's definition \citep[p 243]{H_SA}, there is no guaranteed (strictly positive) profit.%, however there is also a subtle technical point: there is no guarantee that there is no loss if there is an infinite set of outcomes. This is related to the observation that there is no guarantee that an infinite number of monkeys with typewriters will, given enough time,  come up with a work of Shakespeare: it is only that we \emph{expect} them to do so, `almost surely'.  

To understand the connection between the financial concept of arbitrage and the mathematical idea of a martingale measure,  consider  the most basic case   of a single asset whose current price, $X_0$,  can take on one of two (present) values, ${X}_T^D<{X}_T^U$, at time $T>0$, in the future. In this case an arbitrage would exist if $X_0\le X_T^D<X_T^U$: buying the asset now, at a  price that is less than or equal to the future pay-offs, would lead to a possible profit  at the end of the period, with the guarantee of no loss.  Similarly, if ${X}_T^D<{X}_T^U\le X_0$, short selling the asset now, and buying it back  would also lead to an arbitrage. %This is the `Dutch book' argument that Ramsey presented and has been discussed widely in the context of philosophy (for example \citep{H_AFAP} and references therein).
 So, for there to be no arbitrage opportunities we require that
$$
{X}_T^D< X_0 < {X}_T^U.
$$
This implies that there is a  number, $0< q < 1$, such that
\begin{align*}
X_0 =\spa& {X}_T^D + q\,({X}_T^U - {X}_T^D)\\
 =\spa& q\,{X}_T^U +(1-q){X}_T^D.
\end{align*}
The price now, $X_0$, lies between the future prices, $X_T^U$ and $X_T^D$, in the ratio $q:(1-q)$ and represents  some sort of `average'. The first statement of the FTAP can be  interpreted simply as  ``the price of an asset must lie between its maximum and minimum possible (real) future price''.

If $X_0<X_T^D\le X_T^U$  we have that  $q<0$  where as if  $X_T^D\le X_T^U< X_0$ then $q>1$, and in both cases $q$ does not represent a probability measure which by Kolmogorov's axioms,  must lie between 0 and 1.   In either of these cases an arbitrage exists and a trader can make a riskless profit, the market involves \emph{`turpe lucrum'}.  This account gives an insight as to why  James Bernoulli, in his moral approach to probability,  considered situations where probabilities did not sum to 1, he was considering problems that were pathological not because they failed the rules of arithmetic but because they were unfair.  

It follows that if there are no arbitrage opportunities then quantity $q$ can be seen as representing the `probability' that the $X_T^U$ price will materialise in the future.  Formally
\begin{align*}
X_0  =\spa& q\,{X}_T^U +(1-q){X}_T^D\\
\equiv\spa& \e_\q\big[ {X}_T \big].
\end{align*}
The connection between the financial concept of arbitrage and the mathematical object of a martingale is essentially a tautology: both statements mean that the price today of an asset must lie between its future minimum and maximum possible value.

This first statement of the FTAP  was anticipated by Ramsey in 1926 when he defined `probability' in the Pragmatic sense of `a degree of belief' and argues that measuring `degrees of belief' is through betting odds \citep[p 171]{R_TP}.  On this basis he formulates some axioms of probability, including that a probability must lie between 0 and 1 \citep[p 181]{R_TP}.  He then goes on to say that
\begin{quote}
These are  the laws of probability,  \ldots If anyone's mental condition violated these laws, his choice would depend on the precise form in which the options were offered him, which would be absurd. He could have a book made against him by a cunning better and would then stand to lose in any event. \citep[p 182]{R_TP}
\end{quote}

This  is a Pragmatic argument that identifies the absence of the martingale measure with the existence of arbitrage and  today this forms the basis of the standard argument as to why arbitrages do not exist: if they did the, other market participants would bankrupt the agent who was mis-pricing the asset.  This has become known in philosophy as the `Dutch Book' argument and as a consequence of the fact/value dichotomy this  is often presented as a `matter of fact'.  However, ignoring the fact/value dichotomy,  the Dutch book argument is an alternative   of the `Golden Rule'-- ``Do to others as you would have them do to you.''-- it is infused with  the moral concepts of fairness and reciprocity (\cite{W_GR}, \cite{H_AFAP}).  

The essential result of this paper is that embedded at the heart of the first statement of the FTAP is the ethical concept  Justice, capturing the social norms of reciprocity and fairness.  This is significant in the context of Granovetter's discussion of embeddedness in economics \citep{G_EASS}.  It is conventional to assume that mainstream economic theory is `undersocialised': agents are rational calculators seeking to maximise an objective function.  The argument presented here is that a central theorem in contemporary economics, the FTAP, is  deeply embedded in social norms,  despite being presented as an undersocialised mathematical object.  This embeddedness is a consequence of the origins of mathematical probability being in the ethical analysis of commercial contracts: the feudal shackles are still binding this most modern of economic theories. 

Ramsey goes on to make an important  point
\begin{quote}
 Having any definite degree of belief implies a certain measure of consistency, namely willingness to
bet on a given proposition at the same odds for any stake, the stakes being measured in
terms of ultimate values. Having degrees of belief obeying the laws of probability implies a further
measure of consistency, namely such a consistency between the odds acceptable on different
propositions as shall prevent a book being made against you. \citep[p 182--183]{R_TP}
\end{quote}
Ramsey is arguing that an agent needs to employ the same measure in pricing all assets in a market, and this is the key result in contemporary derivative pricing.  Having identified the martingale measure on the basis of a `primal' asset, it is then applied across the market, in particular  to derivatives on the primal asset but the well-known result that if two assets offer different `market prices of risk', an arbitrage exists.  This explains why the market-price of risk appears in the Radon-Nikodym derivative and the Capital Market Line, it enforces Ramsey's consistency in pricing.

The second statement of the FTAP is concerned with incomplete markets, which appear in relation to Arrow-Debreu prices.  In mathematics, in the special case that there are as many, or more, assets in a market as there are possible future, uncertain,  states,  a unique pricing vector can be deduced for the market because of Cramer's Rule.  If the elements of the pricing vector satisfy the axioms of probability, specifically each element is positive and they all sum to one, then the market precludes arbitrage opportunities.  This is the case covered by the first statement of the FTAP. 

In the more realistic situation that there are more possible future states than assets, the market can still be arbitrage free but the pricing vector, the martingale measure, might not be unique.  The agent can still be consistent in selecting which particular martingale measure they choose to use, but another agent might choose a different measure, such that the two do not agree on a price. 
In the context of the Law of One Price, this means that we cannot hedge, replicate or cover, a position in the market, such that  the portfolio  is riskless. The significance of the second statement of the FTAP is  that it  tells us that in the sensible world of imperfect knowledge and transaction costs, a model within the  framework of the FTAP cannot give a precise price. When faced with incompleteness in markets, agents need alternative ways to price assets and behavioural  techniques have come to dominate financial theory.  This feature was realised in  \emph{The Port Royal Logic} when it recognised the role of transaction costs in lotteries.

\section{Two Women and a Duck -- a Pragmatic approach to commerce}\label{s5}

We present the case that the essence of the FTAP is reciprocity, alternatively  Justice and equality in exchange, colloquially fairness.  The pre-history of mathematical probability lies  in Olivi's examination of commercial exchange in the context of Aristotle's \emph{Ethics}. The subsequent emergence of the topic is in the seventeenth century analysis of contracts in the context of `fair' pricing.  In the twentieth century Ramsey provides the `Dutch book' argument, which can be viewed as   the `Golden Rule' of reciprocity.  However, under the influence of a strong fact/value dichotomy that was established in the nineteenth century, the moral injunction not to engage in  \emph{turpe lucrum}, through the practice of arbitrage, becomes highly technical, and ethically neutral, and in the process the essence of reciprocity in the FTAP becomes obscured. 

This argument associates the FTAP with the experimental results of the `Ultimatum Game', an important anomaly for neo-classical economics \citep{T_AUG}.   The game involves two participants and a sum of money.  The first player  proposes how to share the money with the second participant.  The division is made only if the second participant accepts the split,  if the first player's proposal is rejected neither participant receives anything.   The key result is that if the money is not split `fairly' (approximately equally)  then the second player rejects the offer.  This  contradicts the assumption that people are rational utility maximising agents, since if they were the second player would accept any positive payment.  Research has shown that  chimpanzees  are rational maximisers while   the willingness of the second player to accept an offer is dependent on age and culture.  Older  people from societies where exchange plays a significant role are more likely to demand a fairer split of the pot than young children or adults from isolated communities (\cite{MS_UBCA}, \cite{H_FHS}, \cite{H_CPAHS}, \cite{CJM_CRM}). Fair exchange appears to be learnt behaviour developed in a social context and is fundamental to human society and distinguishes the {sapient} member of a \emph{civitas} from the {sentient} animals. The Ultimatum Game provides observational evidence that reciprocity is, or at least should be, a fundamental concept for financial economics.

We have shown the  key role that the FTAP plays in the dominant paradigm of financial economics, involving CAPM (Markowitz portfolio selection), the Efficient Markets Hypothesis (martingales), the use of stochastic calculus and incomplete markets.  At first sight one might assume that this paradigm associated with utility maximisation, but on closer reflection the key components are not.  

Markowitz portfolio theory explicitly observes that portfolio managers are not (expected) utility maximisers, as they diversify,  and offers the hypothesis that a desire for reward is tempered by a fear of uncertainty (\cite{M_PS}, see also \cite[p 432]{R_SF}).  Markowitz's theory was developed into the CAPM by Sharpe while similar models were developed independently by Treynor,  Lintner and  Mossin.  These models conclude that all investors should hold the same portfolio, their individual risk-reward objectives are satisfied by the  weighting of this `index portfolio' in comparison to riskless cash in the bank, a point on the capital market line.  The slope of the CML is the market price of risk, which is  an important parameter in arbitrage arguments.  Significantly, as MacKenzie \citep[pp 86--87]{ENC} observes, Markowitz portfolio selection and CAPM are prescriptive, not descriptive theories; just as medieval merchants were told what was licit by the Scholastics, so,  in the 1980s, asset managers were being told what is `rational' by academics.

Merton had initially attempted to provide an alternative  to Markowitz based on utility maximisation employing stochastic calculus.  He was only able to resolve the problem by employing the hedging arguments of Black and Scholes, and in doing so built a model that was based on the absence of arbitrage, free of \emph{turpe-lucrum}.  The opening paragraph of Black and Scholes includes the prescriptive statement that  ``it should not be possible to make sure profits'', a statement explicit in the Efficient Markets Hypothesis and  in employing an Arrow security in the context of the Law of One Price.

Based on these observations, we conject that the whole paradigm for financial economics, not just the FTAP, is built on the principle of balanced reciprocity.  In order to explore this conjecture we shall examine the relationship between commerce  and themes in Pragmatic philosophy.  Specifically, we  highlight   Robert Brandom's position  that there is
\begin{quote}
a {pragmatist} conception of norms -- a notion of primitive correctnesses of performance {implicit} in {practice} that precludes and are presupposed by their {explicit} formulation in rules and principles. \citep[p 21]{B_MIE}
\end{quote}
The argument that we have presented is  that reciprocity is implicit in the practice of commerce (e.g. \cite{H_BED}) and this norm becomes explicit in Virtue Ethics and then in the early conceptions of mathematical probability.

The `primitive correctnesses' of commercial practices was recognised by Aristotle when he investigated the nature of Justice in the context of commerce and then by Olivi when he looked favourably on merchants.  It is exhibited in the \emph{doux-commerce} thesis, compare  Fourcade and Healey's contemporary description of the thesis
\begin{quote}
Commerce teaches ethics mainly through its communicative dimension, that is, by promoting conversations among equals and exchange between strangers. \citep[p 287]{FH_MVMS}
\end{quote}
with  Putnam's description of Habermas' communicative action  based on 
\begin{quote}
the norm of \emph{sincerity}, the norm of \emph{truth-telling}, and the norm of asserting only what is \emph{rationally warranted} \ldots [and] is contrasted with \emph{manipulation}. \citep[pp 113-114]{P_CFVD}
\end{quote}
There are practices (that should be) implicit in commerce that make it an exemplar of communicative action.

A further expression of markets as centres of communication is manifested in the Asian description of a market as ``Two women and a duck'', which immediately brings to mind Donald Davidson's argument that knowledge is not the product of a bipartite conversations but a tripartite relationship between two speakers and their shared environment (e.g. \cite{D_CTTK}).  The essence of the proverb is that if two women, who are characterised as talkative, and a duck come together, eventually the value of the duck will be determined--knowledge is created.  Replacing the negotiation between market agents with an algorithm that delivers a theoretical price replaces `knowledge', generated through communication, with dogma.  The problem with the performativity that Donald MacKenzie is concerned with \citep{ENC}  is one of monism.  In employing pricing algorithms, the markets cannot perform to something that comes close to `true belief', which can only be identified through communication between  {sapient} humans.  This is an almost trivial observation to (successful) market participants (e.g. \cite{T_FG}, \cite{BS_DR}, \citep[especially Ch 12]{D_HTFRW}), but difficult to appreciate by spectators who seek to attain `objective' knowledge of markets from a distance.

To appreciate the relevance to financial crises of the position that `true belief' is about establishing coherence  through myriad triangulations centred on an asset rather than relying on a theoretical model, consider the comment made by Parliamentary Commission on Banking Standards
\begin{quote}
Excessive complexity in the major banks is not restricted to organisational
structure. The fuelling of the financial crisis by misguided risk models was not simply
the consequence of some mathematicians getting their equations wrong. It was the
result of ignorance, coupled with excessive faith in the application of mathematical
precision, by senior management and by regulators. Many of the elements of this
problem remain. \cite[para. 93, v. II]{PCBS}
\end{quote}
 Mathematicians  understood the limitations of their models, which they communicated.  The problem was that these concerns were not appreciated by policy makers, within an institution, nationally or globally, who appear to have succumbed to the indubitable authority of mathematics \cite[para. 60--61, v. II]{PCBS}.  Stephen Krasner  observes \citep{M_SSSNWIPP} that academics can help policy makers in two respects: ``Provide empirical evidence about what has happened, and offer a conceptual framework through which to understand it.''  A significant issue with the highly technical mathematical models employed in finance is that they lack a ``conceptual framework'' that   non-specialists can understand.  This means that policy makers, whether within or without  banks, cannot ascertain the limitations of mathematical models that inform their decision making. Pragmatism provides the philosophical basis for a conceptual framework that acknowledges both the usefulness and the fallibility of mathematics in finance.

The significance of these issues to the FTAP is captured in a text by Rama Cont and Peter Tankov addressing pricing in markets with discontinuous prices
\begin{quote}
Unless the martingale measure is a by-product of a hedging approach, the price given by such martingale measures is not related to the cost of a hedging strategy therefore the meaning of such `prices' is not clear. \citep[10.5.2]{CT_FMJP}
\end{quote}
If the hedging argument cannot be employed, as in the markets studied by Cont and Tankov, there is no conceptual framework supporting the prices obtained from the FTAP.  This lack of meaning  can be interpreted as a consequence of the strict fact/value dichotomy in contemporary mathematics that came with the eclipse of Poincar\'{e}'s Intuitionism by Hilbert's Formalism and Bourbaki's Rationalism \citep{W_HEBMS}.  The practical problem of supporting the social norms of market exchange has been replaced by a theoretical problem  of developing formal models of markets.  These models then legitimate  the actions of agents in the market without having to make reference to explicitly normative values.

In making this observation and by considering the implications of believing that the FTAP is an expression of reciprocity, we are employing the `Pragmatic maxim' and are making a commitment to real-life experiences.    Another,  more direct, consequence of associating the FTAP with reciprocity is related to the EMH.   Miyazaki observes \citep[p 404]{M_BAS} that speculation by arbitrageurs has been legitimised as ensuring that markets are efficient.  The EMH is based on the axiom that the market price is determined by the balance between supply and demand, and so an increase in trading facilitates the convergence to equilibrium.  If this axiom is replaced by the axiom of reciprocity,  the justification for speculative activity in support of efficient markets disappears.  In fact, the axiom of reciprocity would de-legitimise `true' arbitrage opportunities, as being unfair.  This would not necessarily make the activities of actual market  arbitrageurs illicit, since there are rarely strategies that are without the risk of a loss,  however, it would place more emphasis on the risks of speculation and inhibit the hubris that has been associated with the prelude to the recent Crisis.

These points raise the question of the legitimacy of speculation in the markets.  In an attempt  to understand this issue  Gabrielle and Reuven Brenner identify the  three types of market participant.  `Investors' are preoccupied with future scarcity and so defer income.  Because uncertainty exposes the investor to the risk of loss, investors  wish to minimise uncertainty at the cost of potential profits, this is the basis of classical investment theory. `Gamblers' will bet on an outcome taking odds that have been agreed on by society, such as with a sporting bet or in a casino, and relates to de Moivre's and Montmort's `taming of chance'.    `Speculators' bet on a mis-calculation of the odds quoted by society and the reason  why speculators are regarded as socially questionable is that they have opinions that are explicitly at odds with the consensus: they are practitioners who rebel against a theoretical `Truth' (\citep[p 91]{B_GS},  \citep[p 394] {BS_DR}). This is captured in Arjun Appadurai's argument that the leading agents in modern finance 
\begin{quote}
 believe in their capacity to channel the workings of chance to win in the games dominated by cultures of control \ldots [they] are not those who wish to ``tame chance'' but those who wish to use chance to animate the otherwise deterministic play of risk [quantifiable uncertainty]''. \citep[p 533-534]{A_GFM}
\end{quote}
In the context of Pragmatism, financial speculators embody pluralism, a concept essential to Pragmatic thinking (e.g. \cite{P_MP}, \cite{B_PPHW}, \citep[Ch 2]{B_PT}) and an antidote to the problem of radical uncertainty.

Appadurai was motivated to study finance  by Marcel Mauss' essay \emph{Le Don} (`The Gift'), exploring the moral force behind reciprocity in primitive and archaic societies and goes on to say that  the contemporary financial speculator is  ``betting on the obligation of return'' \citep[p 535]{A_GFM}, and this is the fundamental axiom of contemporary finance.  David Graeber also recognises the fundamental position reciprocity has in finance \citep{G_D}, but where as Appadurai recognises the importance of reciprocity in the presence of uncertainty,  Graeber essentially ignores uncertainty in his analysis that ends with the conclusion that ``we don't `all' have to pay our debts'' \citep[p 391]{G_D}. In advocating that reciprocity need not be honoured, Graeber is not just challenging contemporary capitalism but also the foundations of the \emph{civitas}, based on equality and reciprocity  \citep[p 235]{G_CRI}.

The origins of Graeber's argument are in the first half of the nineteenth century.  In 1836 John Stuart Mill defined political economy as being 
\begin{quote}
  concerned with [man] solely as a being who desires to possess wealth, and who is capable of judging of the comparative efficacy of means for obtaining that end. \citep{M_DPE}
\end{quote}
In \emph{Principles of Political Economy} of 1848 Mill  defended Thomas Malthus'  \emph{An Essay on the Principle of Population}, which focused on scarcity.  Mill was writing at a time when Europe was struck by the Cholera pandemic of 1829--1851 and the famines of 1845--1851 and  while  Lord Tennyson  was describing nature as ``red in tooth and claw''.  At this time, society's fear of uncertainty seems to have been replaced by a fear of scarcity (e.g. \cite{J_WTB}), and these standards of objectivity dominated economic thought through the twentieth century. Almost a hundred years after Mill, Lionel Robbins defined economics as ``the science which studies human behaviour as a relationship between ends and scarce means which have alternative uses''.

Dichotomies emerge in the aftermath of the Cartesian revolution that aims to remove doubt from philosophy \citep[Ch 1]{B_PT}.  Theory and practice, subject and object, facts and values, means and ends are all separated.  In this environment  \emph{ex cathedra} norms, in particular utility (profit) maximisation, encroach on commercial practice.  This is exemplified by the 1950 English court case Buttle v. Sunders ([1950] 2 All ER 193) where it was judged that `my word is my bond' was subordinate to the profit maximisation principle. 

In order to set boundaries on commercial behaviour motivated by profit maximisation, particularly when market uncertainty returned after the Nixon shock of 1971, society imposes regulations on practice. As a consequence, two competing  ethics, functional Consequential ethics guiding market practices and regulatory Deontological ethics attempting stabilise the system, vie for supremacy.  It is in this debilitating competition between two essentially theoretical ethical frameworks that we offer an explanation for the Financial Crisis of 2007-2009: profit maximisation, not speculation, is destabilising in the presence of radical uncertainty and regulation cannot keep up with motivated profit maximisers who can justify their actions through abstract mathematical models that bare little resemblance to actual markets.

This tension is exemplified by the Chartered Financial Analyst (CFA) Institute  \emph{Standards of Practice Handbook} \citep{CFA_SPH}, where the primary obligation is to obey the law, where Buttle v Saunders is tempered by the Basel treaties. There is no discussion of how professionals should interact amongst themselves, only how they interact with clients and employers, agents with whom they have a contractual relationship. This suggests that  a distinction is being  made between the market, populated by analysts, and society as a whole.

An implication of  reorienting financial economics to focus on the markets as centres of `communicative action' is that markets could become self-regulating, in the same way that the legal or medical spheres are self-regulated through professions.  This is not a `libertarian' argument based on  freeing the Consequential ethic from a Deontological brake. Rather it argues that being a market participant entails restricting norms on the agent such as sincerity and truth telling that support  knowledge creation, of asset prices, within a broader objective of social cohesion.  This immediately calls into question the legitimacy of algorithmic/high-frequency trading that seems an anathema in regard to the principles of communicative action.

\section{Conclusion}

The purpose of this paper is to explore the ethical character of contemporary financial economics in light of the Financial Crisis of 2007--2008.  

By examining the contemporary scholarship on the early development of probability we show that the field emerged in the seventeenth century out of the ethical assessment of commercial contracts. In the following century, the \emph{doux-commerce} thesis dominated discussion of the morality of markets, emphasising the role markets play in binding society.  The ethical aspect of probability theory disappears from mathematics at the start of the nineteenth century as science replaces uncertainty with Laplacian determinism \citep{G_EC} and the self-destructive thesis eclipses \emph{doux-commerce}.  Economics developed on Mill's premise that the discipline is ``concerned with [man] solely as a being who desires to possess wealth''  and `value--neutrality' emerges, built on the foundation scientific determinism.  It was within this conceptual framework that the  Black-Scholes equation was developed.  

When a mathematical `theory' to underpin the Black-Scholes-Merton approach, the Fundamental Theorem of Asset Pricing, is developed it relies on Kolmogorov's abstract probabilities.  The essence of this paper is in identifying these `martingale measures' with probabilities that ensure equality in exchange, implicitly imitating the explicitly ethical approach of the early probabilists.  This observation is significant in that it provides evidence of `oversocialisation' in a domain traditionally considered `undersocialised'.

The argument presented in the paper is based on  employing the Pragmatic approach that acknowledges the contingency of knowledge.  By taking this path we argue that markets should be regarded as centres of `communicative action' governed by Pragmatic norms and that recent financial crises have been as a consequence of a dissonance between market participants working to Consequentialist norms but constrained by Deontological norms.  In taking this approach we see a correspondence with Brandom's semantic pragmatism, firstly because we see the implicit norm of reciprocity being made explicit in probability, and secondly because there is a correspondence between the results of the Ultimatum game, which show humans prefer reciprocity to utility maximisation and animals do not, and Brandom's distinction between animal \emph{sentinence} and human \emph{sapience}.  This, in turn, offers a solution to the problems of financial regulation.

An obvious sequel to this paper its to undertake a more detailed and structured investigation into the relationship between commerce and Pragmatism.  This could explore how Pragmatism can inform mathematical developments (such as  in \cite{HH_HUUF}, \cite{MZ_SPBM}, which address means-ends dichotomies, or \cite{BR_DB}, \cite{CW_RFTE}, which address monism/pluralism issues) and how financial intuition can support Pragmatic theories. In particular, a significant topic in Behavioural Finance is Prospect Theory and the transformation of objective probabilities in decision making.  These probability transformations are presented as `irrational' in the context of Empiricism and Realism  but they could be re-interpreted as `warranted' in a Pragmatic framework.  Potentially,  neo-classical and behavioural financial economics could be unified on the basis of reciprocity.

Recent models analysing whether the proliferation of financial instruments leads to instability (e.g. \cite{CMV_EMS} -- which is influential on the widely cited \cite{HM_SRBE}, \cite{S_SRSNFA}) are based on the assumption that agents are seeking to maximise utility, rather than acting in a framework of balanced reciprocity.  An interesting research question would be to investigate what types of financial networks emerge on the basis of either profit maximisation or reciprocity, and then analyse if different network topologies are more resilient to financial shocks.

\section*{Acknowledgements}
I am grateful to Donald MacKenzie and Matthew Festenstein for their formative comments.

\bibliography{/home/timj/LaTex/all}

\end{document}